\documentclass[journal]{IEEEtran}
\IEEEoverridecommandlockouts
\usepackage{comment}
\usepackage[compact]{titlesec}
\titlespacing{\section}{0pt}{*0}{*0}
\titlespacing{\subsection}{0pt}{*0}{*0}
\titlespacing{\subsubsection}{0pt}{*0}{*0}
\usepackage{amsmath,amssymb,amsfonts}
\usepackage{algorithmic}
\usepackage{graphicx}
\usepackage{textcomp}
\usepackage{xcolor}
\usepackage{graphicx}
\usepackage{amsmath}
\usepackage{cite}
\usepackage{wrapfig}
\usepackage{soul}

\usepackage[colorlinks,urlcolor=blue,linkcolor=blue,citecolor=blue]{hyperref}

\usepackage{mdframed}
\usepackage{tcolorbox}
\def\BibTeX{{\rm B\kern-.05em{\sc i\kern-.025em b}\kern-.08em
    T\kern-.1667em\lower.7ex\hbox{E}\kern-.125emX}}

\begin{document}

\title{Knowledge-enhanced Neuro-Symbolic AI for Cybersecurity and Privacy}

\author{\IEEEauthorblockN{Aritran Piplai\IEEEauthorrefmark{1},
Anantaa Kotal\IEEEauthorrefmark{1},
Seyedreza Mohseni\IEEEauthorrefmark{1}, 
Manas Gaur\IEEEauthorrefmark{1}, 
Sudip Mittal\IEEEauthorrefmark{2}, and 
Anupam Joshi\IEEEauthorrefmark{1} \\
\IEEEauthorblockA{\IEEEauthorrefmark{1} Dept. of CSEE, University of Maryland, Baltimore County, Maryland, USA, 21250 } \\
\IEEEauthorblockA{\IEEEauthorrefmark{2} Dept. of CSE, Mississippi State University, Mississippi State, MS, USA, 39762}
}}

\maketitle
\begin{abstract}
    Neuro-Symbolic Artificial Intelligence (AI) is an emerging and quickly advancing field that combines the subsymbolic strengths of (deep) neural networks and explicit, symbolic knowledge contained in knowledge graphs to enhance explainability and safety in AI systems. This approach addresses a key criticism of current generation systems, namely their inability to generate human-understandable explanations for their outcomes and ensure safe behaviors, especially in scenarios with \textit{unknown unknowns} (e.g. cybersecurity, privacy). The integration of neural networks, which excel at exploring complex data spaces, and symbolic knowledge graphs, which represent domain knowledge, allows AI systems to reason, learn, and generalize in a manner understandable to experts. This article describes how applications in  cybersecurity and privacy, two most demanding domains in terms of the need for AI to be explainable while being highly accurate in complex environments, can benefit from Neuro-Symbolic AI.    
\end{abstract}


\section{Introduction and Background}
Neuro-Symbolic AI refers to the integration of neural network-based methods with symbolic knowledge-based approaches. It combines the strengths of both approaches to leverage the representational power of neural networks and the logical reasoning capabilities of symbolic approaches. Neural networks are excellent at large-scale data processing and the extraction of intricate patterns and characteristics from unprocessed input. However, they often struggle to provide explicit explanations for their decisions \cite{1}. This is one of the reasons why, after a promising decade of work from the mid 80s, Neural networks never made it beyond academic and industrial research labs. On the other hand, symbolic knowledge-based approaches, such as rule-based systems or expert systems, utilize explicit knowledge representations and logical reasoning mechanisms. They can capture domain-specific knowledge and provide transparent explanations for their conclusions \cite{1,2}. However, these approaches may struggle with handling uncertain or incomplete information and have limited capacity to learn from large-scale data \cite{1}.

A combination of these two paradigms, called Neuro-Symbolic AI, has started to see popularity among the AI community over the last five years. The idea for this combination is not new, with the term Neural Symbolic being used at least as far back as the early 2000s \cite{2}. In the 90s, there were several efforts to marry connectionist approaches with fuzzy rules, for instance \cite{11}. Indeed, it can (and has) been argued that the kernel of this idea can be found in the McColloch and Pitts' paper titled “A logical calculus of the ideas immanent in nervous activity”. There are multiple reasons for this renewed popularity.  We describe these reasons in the context of cybersecurity.

First, a combination of symbolic reasoning and data-driven methods can be used to extract the sequence of steps or events that triggered the conclusion that the model reached. This is a great motivation for Neuro-Symbolic approaches to be used in cybersecurity and privacy,  especially in solving problems such as threat detection and analysis, which require not just patterns to be detected, but such detected patterns from disparate systems across time to be put into a common context \cite{5}. Neuro-Symbolic approaches can do this while preserving privacy  (e.g. incorporating privacy policies, regulations, compliance).  For example, a Neuro-Symbolic model can reason about sensitive network flow data usage by the neural network detector based on explicit privacy policies and ensure compliance by incorporating privacy-preserving techniques such as differential privacy or secure multi-party computation. 

Second is keeping AI systems secure and safe. The rise in data-driven models for automating vulnerability assessment of systems limits safety, since the system learns only the vulnerabilities it is trained on. In a Neuro-Symbolic approach, the AI-based software systems would be trained with experts acting as adversaries and the AI model would learn to infer policies/rules dynamically \cite{1}. Further, with knowledge in security specification documents explicitly captured using symbolic approaches and used as constraints on behavior, the AI system would be more robust and safe. This is of immediate interest to regulators and legislators in many countries as advanced AI systems have a high probability for generating risky/harmful information, without the control of human knowledge/expertise.

Another reason why a combination of rules and data-driven methods may be useful is the lack of high-quality data to make reliable inferences. This problem can be found in domains where the data for experiments is either hard to obtain, or difficult to share because it may be sensitive. However, alternate sources such as text descriptors of the sensitive data may be available. Rules can be derived from these alternate sources that are shareable. If the available data alone is insufficient to infer reliable conclusions, these rules can be used to augment the conclusions derived from the data \cite{12}. They can also be provided as an input to the data-driven model during the learning process.

Some domains may also be very dynamic and so the data may be representative for only a brief period of time. The conclusions derived from the data may also be valid for a brief period of time. This is a problem in domains such as cybersecurity and fraud-detection. The patterns that we derive from our existing dataset may be useful for some cyber-attacks that are happening right now, but they may not be useful in the future. Combining Deep network based detectors with explicit rules that capture data drift or the temporal limits on the usefulness of a model can help in such situations.

\section{Neuro-Symbolic AI in Cyber}
In the context of cybersecurity, Neuro-Symbolic AI can be applied to enhance various aspects of security systems, such as intrusion detection, malware analysis, vulnerability assessment, and threat intelligence. Very broadly, it can assist in creating the next generation Security Operations Center (SoC) which combines AI approaches with a human either in or on the loop.

Let us consider a scenario of security analysts, who work in a SoC and play a major role in ensuring the security of an organization. The amount of background knowledge they have about evolving and new attacks makes a significant difference in their ability to detect attacks from the output of deep neural networks or machine learning (ML) based systems that today analyze the sensed data stream. We can assist an analyst by capturing information available in open-source threat intelligence sources, such as text descriptions of cyber-attacks or threat feeds, and store them in a structured fashion in a cybersecurity knowledge graph (CKG). We describe two methods in which structured cyber information present in CKGs can be used for downstream tasks, with focus on explainability (reasoning and inference). The first method involves creating sophisticated rules based on real data, and an existing knowledge engine (\textit{rule base framework}). The second method involves using existing rules in downstream data-driven AI models and creating new policies for cybersecurity (\textit{knowledge-guided models}).

In the \textit{rule base framework}, the ultimate goal is to create the strongest and closest rules for target machines to protect them from any type of threats and adversary behaviors. Rules can be simple to complex and they will be consumed by any system or subsystem which needs protection. In the \textit{knowledge-guided models}, we aim to tackle novel cyber-threats or mutated versions of older cyber-threats that do not exist in existing datasets for data-driven experimentation. Exploratory modeling techniques, such as reinforcement learning (RL), are needed to discover new adversaries that can further lead to new defenses. In our experiments, we see that CKGs can guide these exploratory learning strategies to be faster, more effective, and explainable. 

\begin{figure*}[!ht]
    \centering
    \includegraphics[width=0.80\textwidth]{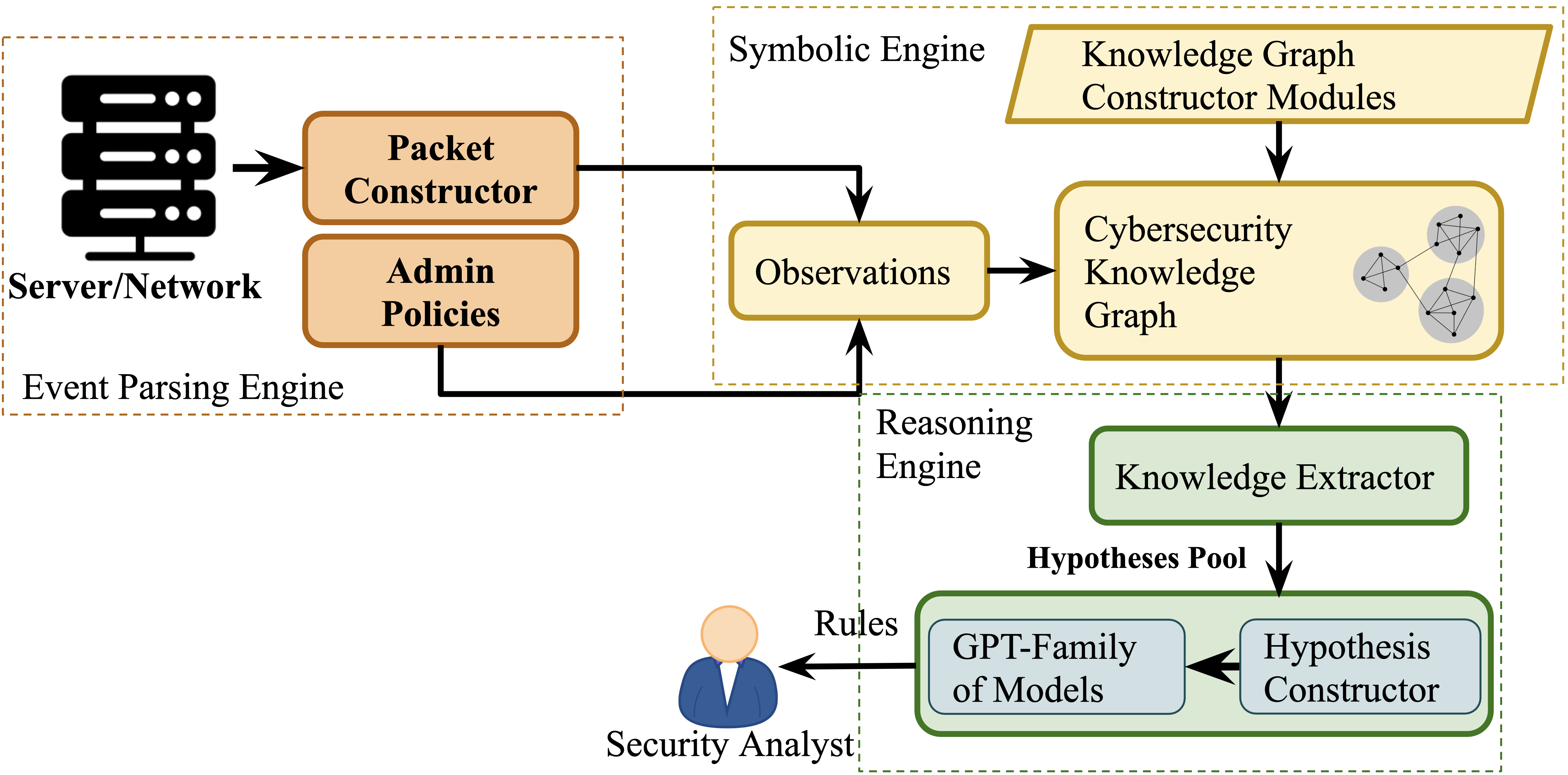}
    \caption{Neuro Symbolic Rule Base framework for dynamic rule inference and generation in cybersecurity}
    \label{fig:arch}
\end{figure*}

\subsection{Modeling Cyber-Events}
A plethora of information is available in unstructured text for cybersecurity. This information can come from various sources such as social media posts, user-written blogs, or published reports from large organizations. Through our research, we were able to extract this unstructured information and convert it into structured knowledge \cite{3}. To achieve this, we utilize semantic triples, which consist of a subject-predicate-object relationship. In other words, when encountering two entities in a text, we aimed to deduce the relationship between them. We use BERT embeddings, as well as neural models, to generate these semantic triples. By analyzing 474 technical reports and numerous smaller technical posts, we construct a knowledge graph (KG) using these semantic triples.

In addition, software companies release information concerning cybersecurity threats to help consumers identify vulnerabilities in their products. This data can be accessed through TAXII (\textit{Trusted Automated eXchange of Intelligence Information}) servers and integrated into a centralized KG.

We model our CKG ontology based on the concepts used in \textit{Structured Threat-Intelligence Exchange} (STIX), an industry standard for exchanging cyber-threat information. We further enrich this ontology with system attribute concepts that describe the malware behavior. The rule base framework and the knowledge-guided models for cybersecurity can use this ontology for generating new rules and policies, and use it in downstream models for explainable intrusion and malware detection \cite{7}. 

\paragraph{Reasoning and Inference Examples:} The semantic triples extracted from open-source text are asserted with CKG. This CKG can be leveraged, not just by human security analysts but also by other data-driven models. We can see some examples of how security analysts can use this CKG to uncover important insights about malware using SPARQL queries seen below.

\begin{minipage}{20em}
\texttt{SELECT ?x where \{
  ?x a FusedCKG:Malware;
     FusedCKG:uses
      FusedCKG:588f41bbc[......].\} }
\end{minipage} \\
\newline
This query asks what malware(s) match a particular hash value.
\newline

\begin{minipage}{20em}
\texttt{SELECT  DISTINCT ?x ?y ?z WHERE \{
    ?x a FusedKG:Malware.
       FusedKG:hasHash [b9d......]
       FusedKG:uses ?y.
    ?y a FusedKG:Attack-Pattern.
    ?z FusedKG:parameterchange 
       FusedKG:increases\_meanchange.\}}
\end{minipage} \\

The query asks what malware has a particular hash and the associated attack pattern for that malware. It also asks what system parameters show an increase when the malware is active. If a framework employs a KG where such dynamic information, in the form of observations, are recorded, then an AI model can leverage such structured knowledge in creating rules when defending against attacks.

\subsection{Rule Base Framework}
The rule base framework is one such architecture that employs existing AI models and transforms them into rule generators. The AI models focus on the knowledge of events, their interlinking with other events or malware in CKG, and a method for generating hypotheses or rules for inference (see Figure \ref{fig:arch}). The three broad sections of this framework are as follows:
\begin{itemize}
    \item \textbf{Event Parsing Engine:} The network block grants access to uncontrolled and unregulated networks. The packets are classified and distinguished to organize and handle data according to their specific types. Administrators, typically network admins, establish administrative policies that contribute to the control and dependability of network traffic. These policies assist in the creation of effective rules for the system.
    \item \textbf{Symbolic Engine:} The Knowledge Graph Constructor serves as a graph builder, while the CKG stores contextual knowledge used to generate hypotheses. The CKG Constructor ensures the provision of valid and accurate context knowledge. The Observation Constructor produces clear and consistent observations derived from network packets and admin policies for integration into the KG. Also, the knowledge test and verification process will be applied to test and verify the alignment (e.g., simply using similarity measures) of both knowledge datasets and observations prior to their incorporation into the CKG.
    \item \textbf{Reasoning Engine:} In a Neuro-Symbolic framework, the Knowledge Extractor component retrieves hypotheses from a KG and combines them with observations to form entailments. These entailments, organized with start and separator tokens, are then processed by a Reasoning Engine composed of transformer-based AI models (e.g., GPT). This engine generates rankings and selects the most suitable hypothesis \cite{4}. The Neuro-Symbolic rule base framework collects and categorizes network packets for use in symbolic engines, generating hypotheses through observation sequences and correlation scores. These hypotheses are combined with the GPT-Family models to create cybersecurity rules, providing improved accuracy, dependability, and explainable behavior in AI malware detection, surpassing traditional rule-based systems like SNORT.
\end{itemize}

\subsection{Rules For Knowledge-guided RL in Cybersecurity}

The semantic knowledge embodied in KGs has the potential to direct algorithms like RL through rewards. While KGs encompass well-established facts regarding cyber incidents, they might not possess details about emerging threats in unpredictable environments. Because of the exploratory nature of RL algorithms, new rules can be inferred based on the static knowledge in KG, systems’ data, and rules extracted from the rule base framework \cite{5}. The reward-based learning in RL allows the intrinsic deep learning AI algorithm to be expressive over the rules and explainable in prediction. 

In the scenario of malware detection, we show that knowledge-guided RL sees even faster convergence, better efficiency, and better response time \cite{5}. The acquired knowledge is incorporated into the exploration phase of the RL algorithm and higher rewards are assigned to states that align with our knowledge sources. Specifically, when we integrate prior knowledge, we observe an 8\% reduction in the average episode time. 
Additionally, we conduct experiments with offline RL algorithms to examine the influence of prior knowledge. We discover that incorporating prior knowledge leads to a 4\% rise in detection in three out of four malware families \cite{6}.

In another study by Piplai et al., we show that employing knowledge-guided reinforcement learning (RL) with rules on extensive Packet Capture (pcap) files, which ranged from 3 to 4 GB in size, leads to more accurate decision-making in countering attackers. In a two-player zero-sum game setup, both the attacker and defender were simulated using a knowledge-guided RL algorithm. By incorporating informed actions based on this approach, we demonstrate a remarkable preservation of 78\% network availability, in comparison to a mere 25\% when knowledge guidance was not utilized \cite{7}.  In both the studies, the knowledge-guided RL strategy would preserve trace to knowledge sources for providing explanations to analysts.

\section{Cybersecurity and Privacy with Knowledge-guided AI and Rule base Framework}
Ensuring privacy in AI models is vital, similar to the importance of cybersecurity. AI models that unintentionally reveal personally identifiable information (PII) during defense against attacks can leave systems vulnerable to future unpredictable attacks. The challenge lies in training AI models to handle various cyber attacks while optimizing response time and preserving privacy. Reliable datasets for privacy-preserving training are scarce, as organizations are hesitant to share data that contains PII or sensitive insights.

To overcome this, generative modeling methods like generative adversarial networks (GANs) can produce surrogate datasets that protect privacy while remaining useful for learning tasks. We utilized conditional GANs (CGANs) along with the t-closeness principle to preserve privacy in tabular data containing continuous and discrete variables \cite{8}. However, training standard CGANs on sensitive data has limitations, as it struggles to model conditionally continuous variables and can only repeat discrete variable values seen in the original dataset.

To address these limitations, we propose a dual approach using privacy-preserving deep learning models, combining generative modeling and symbolic knowledge graphs (KGs) that express domain knowledge. Domain-specific KGs like the Unified Cyber Ontology (UCO) can guide the generative model by providing standardized information \cite{9}. By querying the KG, the CGAN can be trained using a mix of original dataset and discrete values, ensuring that the generated dataset contains observed values and other alternatives from the KG. This approach enhances the privacy preservation in generated datasets for downstream machine learning tasks.

\section{Conclusion and Future Work}
This article discusses two main approaches of Neuro-Symbolic AI in cybersecurity and privacy: (a) the rule base framework and (b) knowledge-guided AI using RL. Both these approaches intrinsically involve partitioning CKG for solving tasks in our concerned domains and ensuring that the AI system is explainable. Our perspective is based on the noticeable improvements over traditional AI systems. However, additional research endeavors are required to develop reasoning engines in cybersecurity and privacy for optimal and explainable decision-making, focusing on the users, such as security analysts. The combination of rules and ML/RL models can further be improved by focusing on which information is more beneficial to the model. The application of transformers, in this case, by selecting specific paths in the graph based on real data is a promising idea. We can also improve this with the help of graph embeddings of state spaces appended to data representation during training time.

Neuro-Symbolic AI has the potential to address critical challenges in the future, especially in domains like privacy in healthcare, where there is a growing demand for explainability. The techniques we explored can be expanded to biomedicine, where medical treatments and procedures are confidential. When limited to specific domains, GANs can only replicate treatment procedures based on the information they were trained on. However, incorporating a Biomedical KG can enhance the GAN model by providing a broader range of knowledge \cite{10}. By delving into such methods of knowledge infusion, we can not only contribute to the cause and effectively tackle the limitations associated with purely data-driven approaches.

\section*{Authors}

\noindent \textbf{Aritran Piplai} is an Assistant Professor at The University of Texas at El Paso (UTEP), starting Fall 2023. He is advised by Prof. Anupam Joshi at The University of Maryland, Baltimore County (UMBC). You can contact him at \texttt{apiplai1@umbc.edu}.

\noindent \textbf{Anantaa Kotal} is a Ph.D. candidate advised by Prof. Anupam Joshi. You can contact her at \texttt{anantak1@umbc.edu}

\noindent \textbf{Seyedreza Mohseni} is a Ph.D. student advised by Prof. Manas Gaur. You can contact him at \texttt{mohseni1@umbc.edu}.  

\noindent \textbf{Manas Gaur} is an assistant professor in the CSEE Department, UMBC. You can contact him at \texttt{manas@umbc.edu}. 

\noindent \textbf{Sudip Mittal} is an assistant professor at the Department of Computer Science \& Engineering at Mississippi State University (MSU). You can contact him at \texttt{mittal@cse.msstate.edu}. 

\noindent \textbf{Anupam Joshi} is the Oros Family Professor in the CSEE Department, and the Acting Dean of the College of Engineering and Information Technology at UMBC. He is also the Director of the UMBC’s Center for Cybersecurity. You can contact him at \texttt{joshi@umbc.edu}.

\bibliographystyle{ieeetr}
\large
\bibliography{Main/references}

\end{document}